\documentclass[osajnl,twocolumn,showpacs,superscriptaddress,9pt]{revtex4-1} 
\usepackage{amsmath,amssymb,graphicx}
\usepackage{times,mathptmx}
\usepackage[colorlinks,linkcolor=red]{hyperref}

\begin{document}

\title{Analysis of high-order dispersion on ultrabroadband microresonator-based frequency combs}

\author{Shaofei Wang}
\affiliation{The Key Lab of Specialty Fiber Optics and Optical Access Network, Shanghai University, 200072 Shanghai, China}

\author{Hairun Guo}
\affiliation{Department of Photonics Engineering, Technical University of Denmark, DK-2800 Kgs. Lyngby, Denmark}

\author{Xuekun Bai}
\affiliation{The Key Lab of Specialty Fiber Optics and Optical Access Network, Shanghai University, 200072 Shanghai, China}


\author{Xianglong Zeng}\email{Corresponding author: zenglong@shu.edu.cn}
\affiliation{The Key Lab of Specialty Fiber Optics and Optical Access Network, Shanghai University, 200072 Shanghai, China}

\begin{abstract}
We numerically investigate the influence of high-order dispersion on both temporal and spectral characterizations of microresonator-based optical frequency combs. The moment method is utilized to study the temporal dynamics of intracavity solitons. The theoretical and numerical results indicate the temporal shifts are induced by high-odd-order dispersion rather than high-even-order dispersion. The role of high-order dispersion on the frequency comb envelopes is carefully elucidated through analyzing the intracavity Cherenkov radiations. We further demonstrate that the spectra envelope of an ultrabroadband optical frequency comb can be engineered by using dispersion profiles with multiple zero dispersion wavelengths.
\end{abstract}


\maketitle

\noindent Frequency combs generated in a monolithic micro-ring resonator have attracted lots of interests soon after its proposal \cite{DelHaye2007,Kippenberg2011}. With short cavity lengths, high fineness and robust platforms, micro-resonators based on Kerr materials, such as $\rm{CaF_2}$ or $\rm{MgF_2}$ crystallines, high-index silica microspheres and silicon nitride micro-rings, can support a frequency comb with high repetition rate, high signal-to-noise ratio and high stabilization. The cavity is pumped with a monochromatic CW laser. Applications include precise frequency metrology, spectroscopy, telecommunications and optical clocks. One promising regime to generate such a frequency comb relies on the four-wave mixing (FWM) parametric process induced by Kerr nonlinearities, also called Kerr frequency comb generation \cite{Matsko2012-PRA}.

The steady state of a frequency comb exhibits mode-locking and the intracavity pulse formation \cite{Matsko2011}, which promises the comb to be highly coherent. Dynamics of the intracavity pulse formation \cite{Lamont2013,Herr2011} and transient regime of the Kerr frequency comb \cite{Savchenkov2012} have been illustrated, which in fact are analogous to pulse propagation dynamics in a Kerr material \cite{Agrawal2006}. The intracavity pulse can form a soliton pattern if the net cavity dispersion is anomalous, or the pulse turns to wave breaking under normal dispersion \cite{Matsko2012-OL}. Significantly, octave-spanning Kerr frequency combs were demonstrated as a result of the ultra-short soliton formation inside the cavity where group velocity dispersion (GVD) is quite weak over a wide wavelength span \cite{Okawachi2011,Zhang2013}. The dispersive wave could also be generated accompanied with the intracavity soliton in the opposite GVD regions, which helps to extend the comb bandwidth just as it does in the well-known supercontinuum generation \cite{Dudley2006}. On the other hand, effects such as pulse self-steepening, material Raman scattering and high-order dispersion (HOD) perturbations may influence the intracavity pulse formation as well as the frequency comb generation, which unfortunately were rarely reported so far.

In this letter we numerically investigate the HOD influence on the intracavity pulse dynamics. Effects of odd-order dispersion (third-order, fifth-order and so on) are revealed to cause temporal delays of the intracavity solitons. Since the HOD effects also work to tailor the cavity dispersion profile with determining one or multiple zero dispersion wavelengths (ZDWs), we present that group-velocity-matched dispersion wave generations \cite{Guo2013} are evoked during the intracavity soliton formation, and Kerr frequency combs are tuned to have different spectral envelopes.

\begin{figure*}[btpb]
  \centering{
  \includegraphics[width = 1 \linewidth]{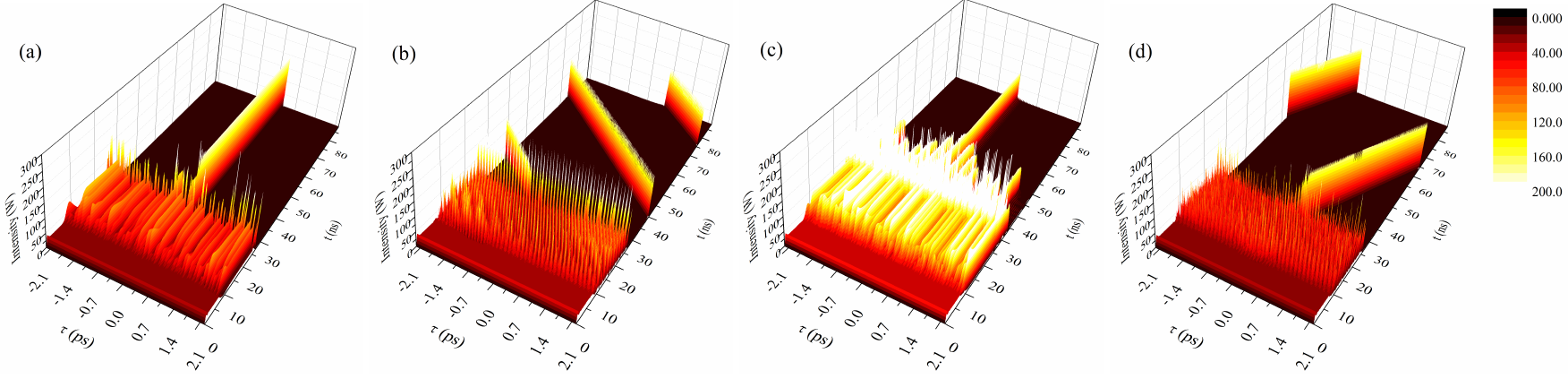}
  }
  \caption{Temporal evolution dynamics of intracavity solitons. (a) only $\beta_2$ ($-2\times{10^{-2}}{\rm{~p}}{{\rm{s}}^2{/\rm m}}$). (b) $\beta_2$ and $\beta_3$ ($-4\times{10^{-4}}{\rm{~p}}{{\rm{s}}^3/{\rm m}}$). (c) $\beta_2$ and $\beta_4$ ($-16\times{10^{-6}}{\rm{~p}}{{\rm{s}}^4/{\rm m}}$). (d) $\beta_2$ and $\beta_5$ ($8\times{10^{-8}}{\rm{~p}}{{\rm{s}}^5/{\rm m}}$).}
    \label{fig1}
\end{figure*}

The numerical model is the Lugiato-Lefever (LL) equation \cite{Haelterman1992,Coen2013} which is actually a periodically damped nonlinear Schr\textbf{$\ddot{\rm o}$}dinger equation driven by an external pump source, describing the intracavity pulse dynamics

\begin{align}
{t_R}\frac{{\partial E\left( {t,\tau } \right)}}{{\partial t}} = \left[ { - \alpha  - i{\delta _0} + iL\sum\limits_{k \ge 2} {\frac{{{\beta _k}}}{{k!}}{{\left( {i\frac{\partial }{{\partial \tau }}} \right)}^k}} } \right. \notag \\
\left. { + i\gamma L{{\left| E \right|}^2}} \right]E\left( {t,\tau } \right) + \sqrt \theta  {E_{in}},
\label{eq1}
\end{align}
where $E\left( {t,\tau } \right)$ is the intracavity electric field. $t$ denotes the slow time with respect to consecutive roundtrips. $\tau$ is a fast time defined in a reference frame traveling at the group velocity of $E$. $t_R$ is the round-trip time relate to the cavity length $L$. $\alpha$ represents the linear cavity loss, $\theta$ is the transmission coefficient, and the critical coupling state occurs when $\alpha = \theta$. The phase detuning parameter ${\delta _0}$ is chosen by scanning the CW laser wavelength. ${\beta _k}$ accounts for the HOD coefficients at the central frequency of the driving field. $\gamma$ is the nonlinear coefficient.

The moment method is an extensively used approach to describe pulse dynamics in optical fibers governed by the nonlinear Schr\textbf{$\ddot{\rm o}$}dinger equation \cite{Grigoryan1999,Santhanama2003}. Since Eq. (\ref{eq1}) is formally similar to the Schr\textbf{$\ddot{\rm o}$}dinger equation, we herein also use this method to analyze the comb dynamics. For an optical resonator, the dynamics of final steady-state pulses are mainly focused because the initial evolution stage is fairly chaotic \cite{Matsko2013}. In the moment method,pulse energy ($E_p$), temporal position ($T_p$) and frequency shift ($\Omega _p$) are calculated with the following set of moments

\begin{align}
&{E_p} = \int_{ - \infty }^{ + \infty } {{{\left| E \right|}^2}d\tau }, {T_p} = \frac{1}{{{E_p}}}\int_{ - \infty }^{ + \infty } {\tau {{\left| E \right|}^2}d\tau }, \label{eq2} \\
&{\Omega _p} = \frac{i}{{2{E_p}}}\int_{ - \infty }^{ + \infty } {\left( {{E^*}\frac{{\partial E}}{{\partial \tau }} - E\frac{{\partial {E^*}}}{{\partial \tau }}} \right)} d\tau. \label{eq3}
\end{align}

\noindent Substituting Eqs. (\ref{eq2}) and (\ref{eq3}) into Eq. (\ref{eq1}) and after some algebra, we get

\begin{align}
&\frac{{d{E_p}}}{{dt}} = 0, \frac{{d{\Omega _p}}}{{dt}} = 0,  \label{eq4} \\
&\frac{{d{T_p}}}{{dt}} = \sum\limits_{k \ge 1} {\left( {\int_{ - \infty }^{ + \infty } {\frac{{L{\beta _{2k + 1}}}}{{\left( {2k + 1} \right)!}}{{\left| {\frac{{{\partial ^k}E}}{{\partial \tau }}} \right|}^{2k}}d\tau } } \right)}  + \notag \\
&{\rm{~~~~~~~~~~}}\sum\limits_{k \ge 1} {\left( {\int_{ - \infty }^{ + \infty } {\frac{{iL{\beta _{2k}}}}{{\left( {2k} \right)!}}\left( {{\Theta _k} \left( {t,\tau } \right)} \right)d\tau } } \right)}, \label{eq5}
\end{align}
where

\begin{align}
{\Theta _k} \left( {t,\tau } \right) = \frac{{{\partial ^{k - 1}}{E^*}}}{{\partial \tau }}\frac{{{\partial ^k}E}}{{\partial \tau }} - \frac{{{\partial ^{k - 1}}E}}{{\partial \tau }}\frac{{{\partial ^k}{E^*}}}{{\partial \tau }}.
\label{eq6}
\end{align}

\noindent Evidently, the frequency shift with respect to the slow time $t$ should be a constant from Eq. (\ref{eq4}). Hence, $\Omega _p = 0$ must be fulfilled for a stable frequency comb generation with fixed frequencies. Furthermore, using the following relations

\begin{align}
&{\mathcal{F}}\left[ {\frac{{{\partial ^k}E\left( {t,\tau } \right)}}{{\partial {\tau ^k}}}} \right] = {\left( { - i\omega } \right)^k}\tilde E\left( {t,\omega } \right),k = 1,2,3, \ldots.\, \label{eq7} \\
&\frac{{\partial \left[ {{{\left( { - i\omega } \right)}^k}\tilde E\left( {t,\omega } \right)} \right]}}{{\partial t}} = 0, \label{eq8}
\end{align}
the frequency shift of the k-th order derivation of $E\left( {t,\tau} \right)$ should be zero as well, therefore yields ${\Theta _k}\left( {t,\tau } \right) = 0$. Eventually, Eq. (\ref{eq5}) can be reduced to

\begin{align}
\frac{{d{T_p}}}{{dt}} = \sum\limits_{k \ge 1} {\left( {\int_{ - \infty }^{ + \infty } {\frac{{L{\beta _{2k + 1}}}}{{\left( {2k + 1} \right)!}}{{\left| {\frac{{{\partial ^k}E}}{{\partial \tau }}} \right|}^{2k}}d\tau } } \right)}.
\label{eq9}
\end{align}

\noindent Equation (\ref{eq9}) clearly indicates that the temporal shifts of intracavity solitons are exclusively determined by high-odd-order dispersion (HOOD) rather than high-even-order dispersion (HEOD). In order to validate the above theoretical predictions, we numerically solve Eq. (\ref{eq1}) by exploiting the split-step Fourier method. The resonator parameters used here are same with that of a recently common used silicon nitride microresonator \cite{Okawachi2011,Coen2013}. Typical simulation results are shown in Fig. \ref{fig1}, in which the resonators are all operated in the anomalous GVD regime, and the HOD terms are expanded up to five order. Due to the intracavity initial modulation instability, the incident field firstly undergoes a chaotic period, then periodic pulse series are generated, and a Kerr frequency comb with arbitrary phase relation is formed in frequency domain \cite{Coen2013U}. If we suitably tune the parameter $\delta_0$, the overall field will evolve towards the final stage consisting of a single soliton superimposed upon a weak CW background. As a consequence, the steady-state frequency comb is formed with definite phase relation, however, from the perspective of temporal domain, the intracavity solitons continuously drift along the fast time axis $\tau$ due to the HOD influence. Specifically, if we consider only GVD term in Fig. \ref{fig1}(a) or GVD together with $\beta_4$ in Fig. \ref{fig1}(c), the solitons evolve without any temporal drift. Whereas the temporal shifts can be clearly observed when the HOODs are taken into account, as shown in Fig. \ref{fig1}(b) ($\beta_2$ and $\beta_3$) and Fig. \ref{fig1}(d) ($\beta_2$ and $\beta_5$). These features are well understood from the theoretical predictions.

\begin{figure}[hbtp]
  \centering{
  \includegraphics[width = 1 \linewidth]{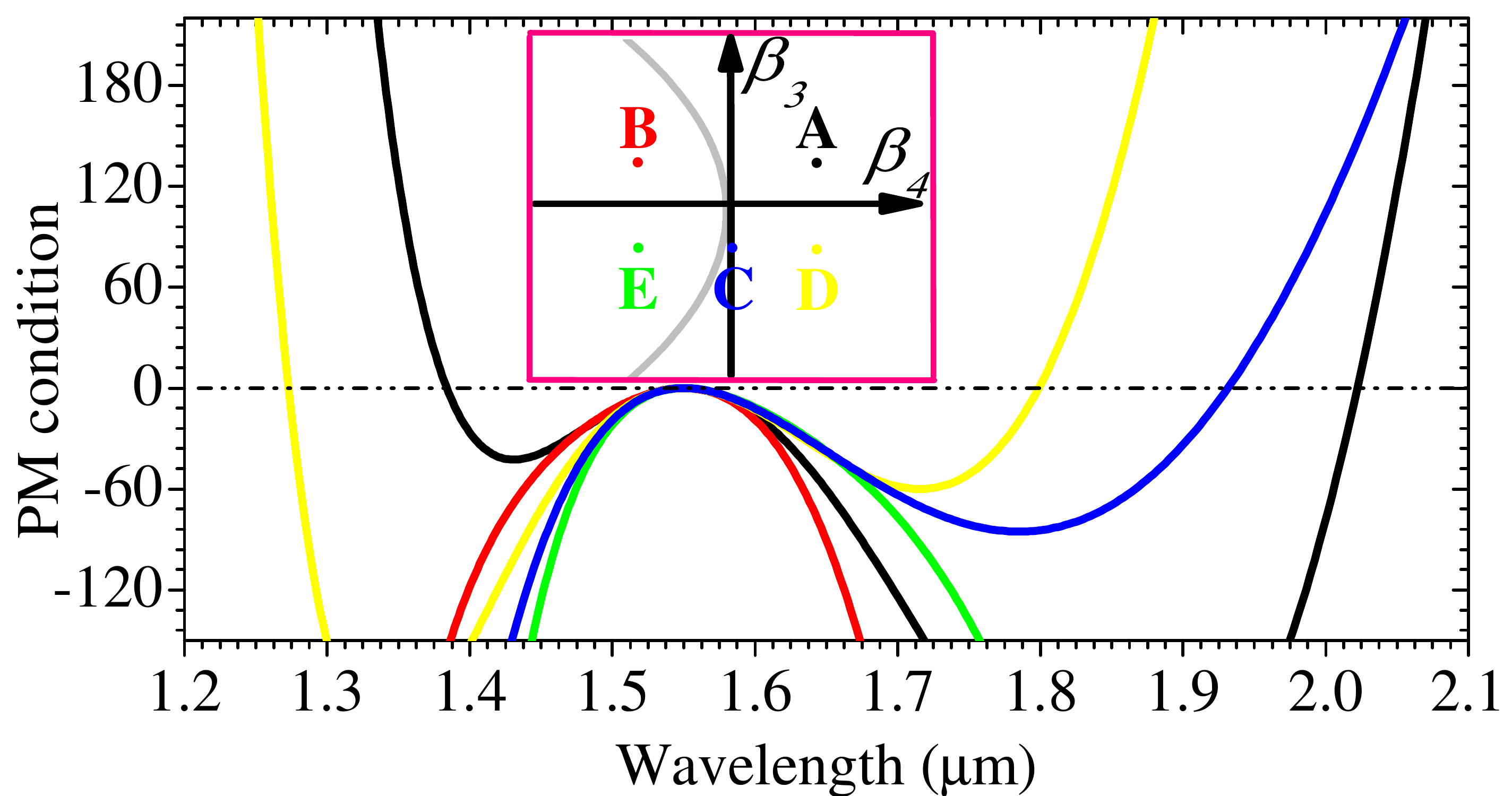}%
  }
  \caption{Schematic PM relations in different regions in ($\beta_3$, $\beta_4$) space. The gray line in the inset denotes the boundary of different PM features. $\beta_2$ is $-2\times{10^{-2}}{\rm{~p}}{{\rm{s}}^2/{\rm m}}$; $|\beta_3|$, $|\beta_4|$ are $4\times{10^{-4}}{\rm{~p}}{{\rm{s}}^3/{\rm m}}$ and $8\times{10^{-6}}{\rm{~p}}{{\rm{s}}^4/{\rm m}}$, respectively. A, B, D and E locate in ($\beta_3$, $\beta_4$) with different signs and C is ($\beta_3$, 0). All PM lines in different colors are corresponding with the colors of A, B, C, D, E in the inset.}
  \label{fig2}
\end{figure}

Next we focus the HOD influence on the envelopes of Kerr frequency combs. The ubiquitous phenomenon behind the HOD effect is dispersive wave generation, also called optical Cherenkov radiation (OCR), resulting in frequency-shifting spectra generated from solitons \cite{Akhmediev1995,Tsoy2007,Roy2009,Herr2012A}. As reported in \cite{Coen2013}, OCR can also be achieved in microresonators when the phase matching (PM) is satisfied between the intracavity soliton and the linear radiation. The PM condition becomes ${\rm PM}\left( \Omega  \right) = \sum\limits_{k = 2}^4 {\frac{{{\beta _k}\left( {{\omega _0}} \right)}}{{k!}}{{\left( {\omega - {\omega _0}} \right)}^k}}$ with a negligible soliton wave number, i.e., only considering $\beta_2$, $\beta_3$ and $\beta_4$ and removing the nonlinear term. Here $\Omega$ is the frequency difference between the pump frequency $\omega$ and the frequency of dispersive wave $\omega_0$. Actually, the possible wavelengths of the OCRs are the real roots satisfying ${\rm PM} = 0$. The generated PM topology shown in the inset of Fig. \ref{fig2} indicating that two different regions are divided by the gray line in ($\beta_3$, $\beta_4$) plane. Specifically, on the right side of the gray line, single or double real roots exist, such as A, C and D, while no zero PM wavelength exists on the left side, such as points of B and E. Their corresponding colored PM lines are depicted in Fig. \ref{fig2}.

\begin{figure}[htbp]
  \centering{
  \includegraphics[width = 1 \linewidth]{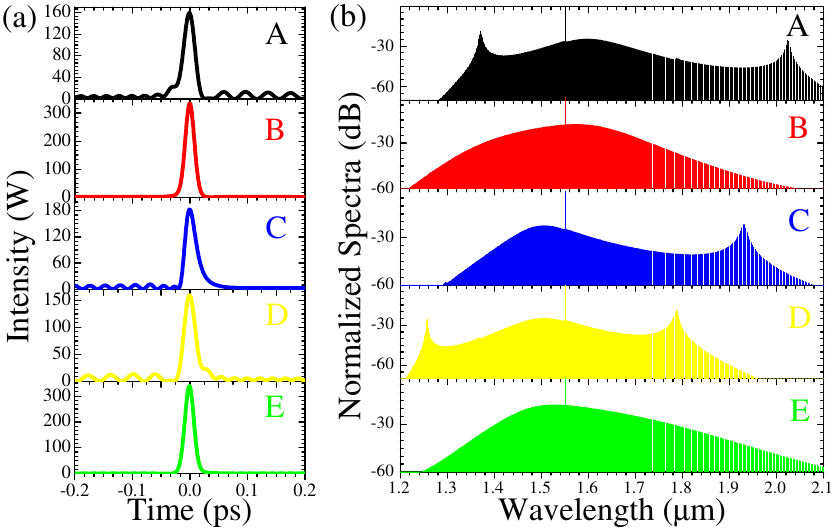}%
  }
  \caption{Final steady-state (a) temporal and (b) spectral envelopes calculated based on same-color dispersion curves shown in Fig. \ref{fig2}.}
  \label{fig3}
\end{figure}

By solving Eq. \ref{eq1} employing the dispersion parameters in Fig. \ref{fig2}, the stable frequency combs and intracavity solitons are obtained, as shown in Fig. \ref{fig3}. Obviously, the smooth temporal and spectral envelopes can be seen in the cases of B and E in Fig. \ref{fig2} due to no existence of OCRs. Instead, For the dispersion curves of A, C and D in Fig. \ref{fig2}, both the complex spectra and temporal sidebands arise. Either advance or delay of a sideband in the temporal domain is governed by the perturbations of the generated OCRs to the intracavity soliton. Double sideband oscillations appear for B and D and single oscillation can be found in C due to the PM conditions.

On the other hand, the opposite sign of $\beta_3$ (e.g., from A (B) to D (E) in Fig. \ref{fig3}) leads to the mirror-symmetry envelopes in both temporal and spectral domains. Nevertheless, comparing A (D) with B (E), we find that the sign of $\beta_4$ dominates the switching of the OCRs. Further expanding the dispersion to higher orders are introduced in \cite{Roy2009}, they found that the HEODs cause either a pair of conjugate OCRs or no OCR at all, whilst the HOODs merely lead to a single OCR. As a result, the HEODs unlikely induce any asymmetry to the temporal profile of a soliton, but the HOODs are possible. The asymmetry of temporal pulse shape leads to a temporal shift of the intracavity solitons, which is same behaviors with an asymmetric Airy-shaped pulse accelerating during propagation \cite{Siviloglou2007}. It should be noted here that all of the final intracavity solitons shift somehow along $\tau$ axis, but all of them are shifted to the peak positions for a well comparison of the envelopes in Fig. \ref{fig3}(a).

\begin{figure}[hbtp]
  \centering{
  \includegraphics[width = 1 \linewidth]{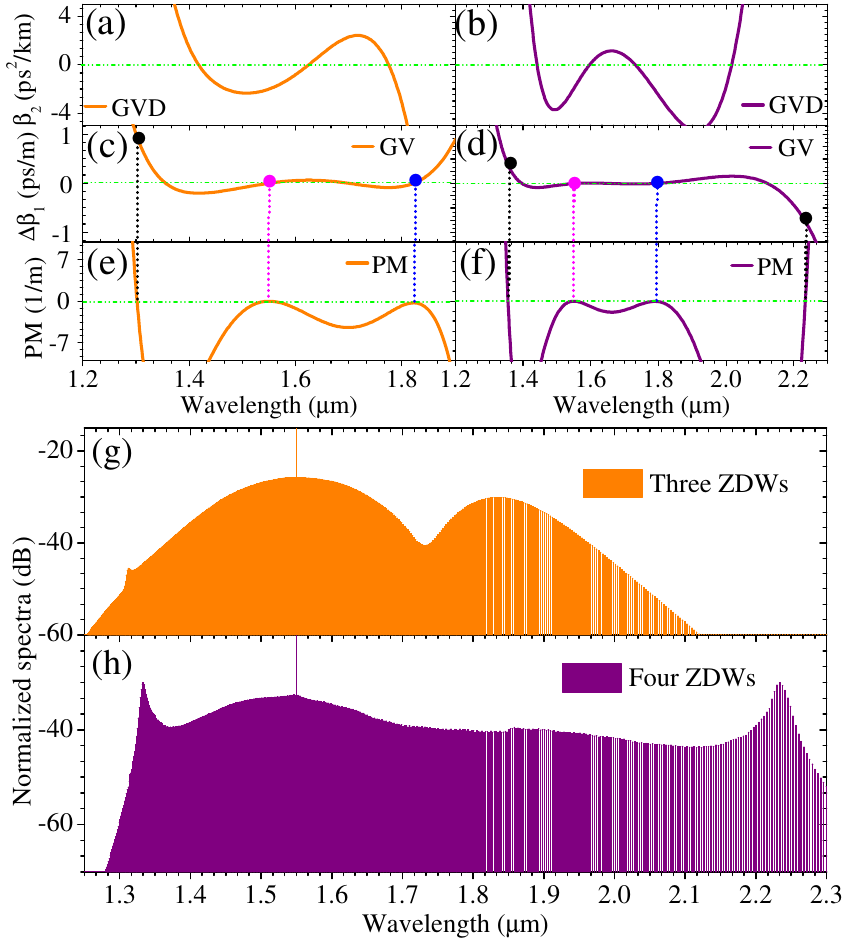}
  }
  \caption{Engineering the spectra of ultrabroadband frequency combs by tuning the GVD profiles with (a) three and (b) four ZDWs. The corresponding group velocity profiles are (c) and (d) as well as the PM conditions in (e) and (f), respectively. (g) and (h) are the resultant comb spectra.}
  \label{fig4}
\end{figure}

Furthermore, the spectrum of an ultrabroadband frequency comb can be engineered by suitably tailoring the dispersion profiles if including more HOD terms. Two typical GVD profiles with multiple ZDWs are presented here due to their rich PM relations, as shown in Figs. \ref{fig4}(a) and \ref{fig4}(b). Note here the HOD terms are expanded up to 10-th order. Three and four ZDWs can be found due to the strong waveguide dispersion counterbalancing the material dispersion, as seen in Ref. \cite{Zhang2013,Wang2013}.

The group-velocity (GV) matching between the soliton and the radiation waves are given in Figs. \ref{fig4}(c) and \ref{fig4}(d) and their corresponding PM curves in Figs. \ref{fig4}(e) and \ref{fig4}(f), respectively. For a GVD profile with three ZDWs, the GV of the intracavity soliton (i.e., pink point) is equal to that of the OCR at longer wavelength (i.e., blue point), while it is faster than another OCR at shorter wavelength (i.e, black point). The GV matching between the soliton and the OCR allows sufficient interaction between the soliton and the OCR, leading to a class of broadband, highly efficient OCR, as shown in Fig. \ref{fig4}(g). The OCR at longer wavelength is broader, more efficient than that at shorter wavelength. Similar features are found in case of a GVD profile exhibiting four ZDWs, in which a GV-matching OCR promotes a wide range of spectra between the pump wavelength and narrow band OCR beyond $2.2~\mu m$. In both cases, considerable energy from the first anomalous GVD range is transferred into the second anomalous GVD region at longer wavelength, so that the overall comb envelopes expand to even an octave spectral range with high flatness.

In summary, the HOD influence on both temporal intracavity solitons and spectral characterizations of the ultrabroadband microresonator-based optical frequency combs have been studied. We find that the temporal shifts of the intracavity solitons are merely determined by the HOOD terms, which are well understood by exploiting the moment method. Roles of the HOD terms in affecting the envelopes of the Kerr frequency combs are also analyzed by explaining the intracavity OCRs. Finally, two typical GVD profiles with multiple ZDWs by tuning the HODs are given to engineer the overall frequency envelopes of Kerr combs. Our analysis will give insight for future exploring spectral properties of ultrabroadband optical frequency combs.

This work was supported in part by National Natural Science Foundation of China (11274224), Shanghai Shuguang Program (10SG38) and the open program (2013GZKF031307) from State Key Laboratory of Advanced Optical Communication Systems and Networks, Shanghai Jiao Tong University. Xianglong Zeng acknowledges the Research support by the Program for Professor of Special Appointment (Eastern Scholar) at Shanghai Institutions of Higher Learning.


\end{document}